\documentstyle[12pt,aaspp4]{article}

\lefthead{Intermittency Spectra in Active Regions}
\righthead{Abramenko}

\begin{document}

\title{Intermittency Spectra of the Magnetic Field in Solar Active
Regions}

\author{Valentyna Abramenko and Vasyl Yurchyshyn}
\affil{Big Bear Solar Observatory, 40386 N. Shore Lane, Big Bear City, CA
92314}

\begin{abstract}

We present results of a study of intermittency and multifractality of magnetic
structures in solar active regions (ARs). Line-of-sight magnetograms for 214 ARs
of different flare productivity observed at the center of the solar disk from
January 1997 until December 2006 are utilized. Data from the Michelson Doppler
Imager (MDI) instrument on-board the {\it Solar and Heliospheric Observatory}
(SOHO) operating in the high resolution mode, the Big Bear Solar Observatory
digital magnetograph and {\it Hinode} SOT/SP instrument were used. Intermittency
spectra were derived via high-order structure functions and flatness functions.
The flatness function exponent is a measure of the degree of intermittency. We
found that the flatness function exponent at scales below approximately 10 Mm is
correlated to the flare productivity (the correlation coefficient is - 0.63).
{\it Hinode} data show that the intermittency regime is extended toward the
small scales (below 2 Mm) as compared to the MDI data. The spectra of
multifractality, derived from the structure functions and flatness functions,
are found to be more broad for ARs of highest flare productivity as compared to
that of low flare productivity. The magnetic structure of high-flaring ARs
consists of a voluminous set of monofractals, and this set is much richer than
that for low-flaring ARs. The results indicate relevance of the multifractal
organization of the photospheric magnetic fields to the flaring activity. Strong
intermittency observed in complex and high-flaring ARs is a hint that we observe
a photospheric imprint of enhanced sub-photospheric dynamics.

\end{abstract}


\section { Introduction}

Energy release processes in solar active regions (ARs) could be, in general,
divided into two types. The first one is a quasi-stationary energy release,
which is responsible, for example, for heating of ARs and manifests itself as
continuous soft X-ray and ultra-violet emissions above an AR. It is well known
that the total magnetic flux and the area of an AR are very well correlated with
the total soft X-ray luminosity (Fisher at al. 1998). Contributions from
adjacent domains of an AR are summed in the total X-ray luminosity. In this
sense, the process might be considered as an additive, i.e., a linear process.
Forecasting of such a process is straightforward and reliable.

Another type of energy release is an impulsive process known as solar flares of
various energy scales: from nano-flares ($10^{24}$ ergs) to super-strong
eruptions of $10^{33}$ ergs (see, e.g., Shimizu \& Tsuneta 1997). A drastically
different situation is encountered here. The physical background for the flare
origin is a non-linear process occurring in a large volume ranging from the
convective zone to the corona. Non-linear processes are multiplicative in nature
and involve inter-scale energy exchange, in other words, a non-linear energy
cascade. This is a way of evolution of any turbulent medium. Large fluctuations
in temporal and spatial domains are nor rare and they contribute significantly
into the main values. The temporal and spatial images of the system (an AR in
our case) acquire an intermittent (or, in other words, multifractal) character.
Strictly speaking, in such a system, it is impossible to exactly predict the
location and time of next large fluctuation (a solar flare in our case).
However, since solar eruptions play a key role in Earth space weather, numerous
empirical approaches were proposed during the last two decades to forecast solar
flaring activity (see, e.g., the recent reviews by McAteer et al. 2009 and
Georgoulis 2010 and references in). 

In the present study, we do not offer yet another approach for solar flare
prediction. We will rather focus on non-linearity and complexity as a function
of spatial scales in solar magnetic fields - an environment hosting flares and,
in many senses, responsible for flare activity in ARs. It is well known that a
non-linear system can not be characterized adequately by only one scalar
parameter. To explore the inter-scale energy exchange, we should consider
various spectral functions of the system. Therefore, we focus here on
intermittency and multifractality spectra of the photospheric magnetic field in
various active regions displaying different flare productivity.

\section { Data }

A set of line-of-sight magnetograms, recorded for 214 ARs from January 1997
until December 2006, was analyzed. The bulk of the magnetograms (212) were
obtained with the SOHO/MDI instrument (Scherrer et al. 1995) in the high
resolution mode (MDI/HR, the pixel size of 0.6 $^{''}$). Data for two ARs (NOAA
9393 and 10720) were obtained with the Big Bear Solar Observatory's Digital
Magnetograph (BBSO/DMG, Spirock 2005) at very good seeing conditions (pixel size
of 0.6 $^{''}$). For NOAA AR 10930, we also utilized {\it Hinode} line-of-sight
magnetograms derived with the SOT/SP instrument (Tsuneta et al. 2008). In all of
the cases, the region of interest was located near the solar disk center (no
farther than 20$^\circ$  away from the central meridian), so that the projection
effect was negligible.

ARs selection was performed for this study. First, only those ARs which were
located inside the MDI/HR field-of-view (FOV) were analyzed. This requirement
results in excluding of a large number of high-latitude ARs. And gaps in the
MDI/HR data coverage diminished out data set. Second, a threshold on the total
flux of an AR in question should be specified, otherwise there is no limit where
to stop. For example, ephemeral active regions (e.g., Harvey \& Martin 1972,
total flux of about 10$^{20}$ Mx) are also active regions, although not
associated with strong flaring. Magnetic structures of small compact ARs are
usually purely resolved by MDI. Thus, to be consistent and analyze all ARs with
the compatible quality, we have to select only ARs with total flux exceeding
some threshold value. As a good compromise between the quality of analyzed data
and the data set size, we chose this threshold value as 10$^{22}$ Mx. The total
flux was calculated as a sum of absolute values of flux densities in all pixels
of the magnetogram. For each AR, we analyzed one magnetogram.

The flare productivity of an AR was measured by the flare index, $A$, introduced
in Abramenko (2005a). Since the X-ray classification of solar flares (X, M, C,
and B) is based on denary logarithmic scale, we can define the flare index as 
\begin{equation}
A=(100 S^{(X)}+10 S^{(M)}+ S^{(C)}+0.1 S^{(B)})/t.
\label{A}
\end{equation}
Here, $S^{(j)}$ is the sum of all GOES flare magnitudes of a certain X-ray
class:
\begin{equation}
S^{(j)}=\sum_{i=1}^{N_j} I_i^{(j)},
\label{AA}
\end{equation}
where $N_j=N_X, N_M, N_C$ and $N_B$ are the numbers of flares of X, M, C and B
classes, respectively, that occurred in a given active region during its passage
across the solar disk that is represented by the time interval $t$ measured in
days. $I_i^{(j)}=I_i^{(X)}, I_i^{(M)}, I_i^{(C)}$ and $I_i^{(B)}$ are GOES
magnitudes of X, M, C and B flares. Interval $t$ was taken to be 27/2 days for
the majority of the ARs with exception of emerging ones. In general, those ARs
that produced only B and C-class flares have the flare index smaller than 2,
whereas several X-class flares will result in the flare index exceeding 100.
Zero flare index does not indicate the absence of any flares. It rather means
that our tools are not capable to detect extremely weak flares for a given AR.
Again, in view of a question "where to stop", we chose to restrict ourselves by
ARs of non-zero flare index as it was measured from the GOES data on the basis
of Eqs. (\ref{A} - \ref{AA}). Therefore, the flare index, $A$, analyzed here,
represents an average rate of the flare productivity of an ARs during the time
interval of the AR's presence on the solar disk, but not the probability of
imminent flare.

Thus, out data set by no means is a complete one. It rather includes ARs in the
tail of an ARs distribution (e.g., Zhang 2010) - a subset of ARs with moderate
and large total flux and non-zero flare productivity. However, it is exactly
these kind of events that define the space weather.

We note that, to some extent, the flare index may be considered to be a measure
of intermittency in the {\it time} domain. Indeed, each flare in the time
profile of integrated solar X-ray emission represents a strong fluctuation above
the nearly continuous background of much weaker fluctuations. This description
is in a perfect agreement with the definition of intermittency. Along with this,
more rigorous measures for the temporal intermittency for ARs were suggested
(see, e.g., Abramenko et al. 2008).

\section { Intermittency Spectra: Method }

There are many approaches to probe intermittency and multifractality (see, e.g.,
recent review by McAteer et al. 2009 for applications to the solar magnetic
fields). We traditionally utilize the high-order structure functions approach
(Stolovitzky \& Sreenivasan 1993; Frisch 1995; Consolini et al. 1999; Abramenko
et al. 2002, 2003, 2008; Abramenko 2005b, 2008; Georgoulis 2005; Buchlin et al.
2006; Uritsky et al. 2007). 

\begin{figure}[!h] \centerline {\epsfxsize=5.0truein
\epsffile{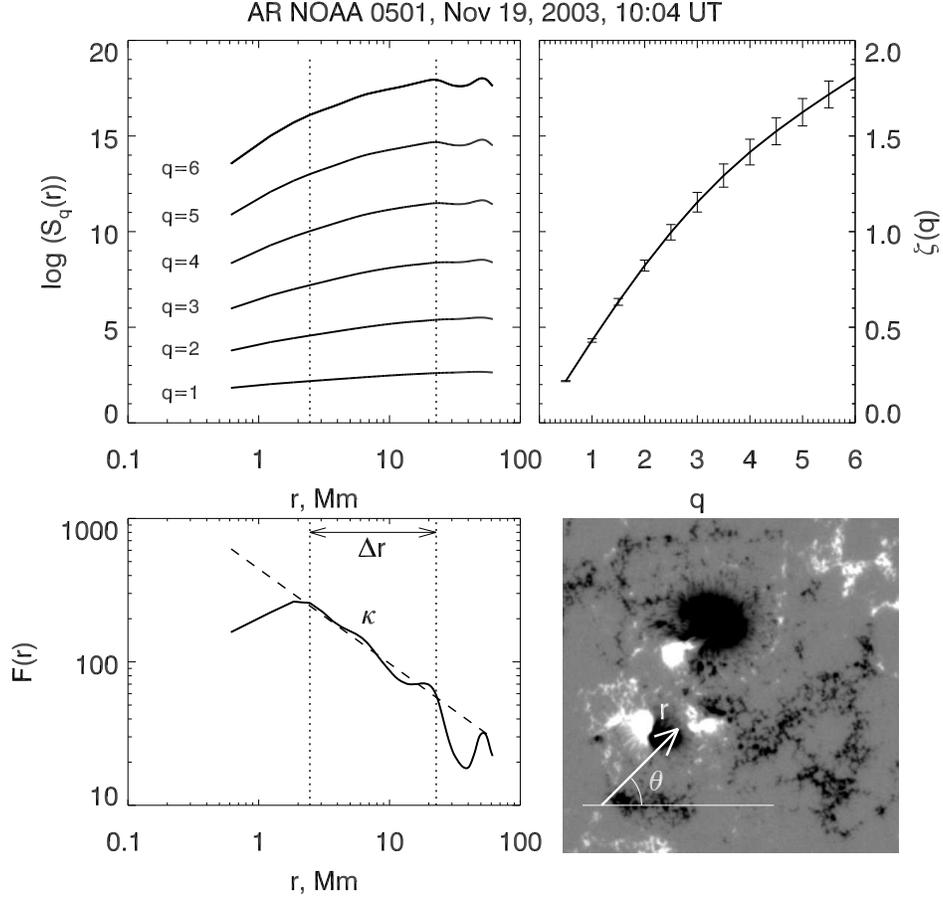}}
\caption{\sf Structure functions $S_q(r)$ ({\it upper left}) calculated from a
magnetogram of active region NOAA 0501 ({\it lower right}) according to Equation
(\ref{Sq}). {\it Lower left} - flatness function $F(r)$ derived from the
structure functions using Equation (\ref{Fr}). Error bars for $S_q(r)$ and
$F(r)$ are no larger than the thickness of the lines, due to large statistics in
Eq. \ref{Sq}. Vertical dotted lines in both left panels mark the interval of
intermittency, $\Delta r$, where flatness grows as power law when $r$ decreases.
The index $\kappa$ is the power index of the flatness function determined within
$\Delta r$. The slopes of $S_q(r)$, defined for each $q$ within $\Delta r$,
constitute $\zeta(q)$ function ({\it upper right}), which is concave (straight)
for an intermittent (non-intermittent) field. An example of a separation vector
$\bf{ r}$ and the corresponding directional angle $\theta$ between the positive
direction of the $x$-axis (EW-axis) and the vector $\bf{ r}$ (see Section 5 for
further discussion) are shown on the magnetogram.}
\label{fig1} 
\end{figure}

Structure functions were introduced by Kolmogorov (1941) (the original Russian
paper was later republished as Kolmogorov 1991) and are defined as statistical
moments of the $q-$powers of the increment of a field. In our case, the analyzed
field is the LOS component of the photospheric magnetic field, $B_l$ (see Figure
\ref{fig1} for illustration), so the corresponding structure function can be
written as 
\begin{equation}
S_q(r) = \langle | {B_l}({\bf x} + {\bf r}) - {B_l}({\bf x})|^q \rangle, 
\label{Sq} 
\end{equation}  
where ${\bf x}$ is the current pixel on a magnetogram,  ${\bf r}$ is the
separation vector between any two points used to measure the increment (see the
lower right panel in Figure \ref{fig1}), and $q$ is the order of a statistical
moment, which takes on real values. Here angular brackets denote averaging over
the magnetogram, and the vector $\bf r$ is allowed to adopt all possible
orientations, $\theta$, from 0$^\circ$ to 180$^\circ$ (because of the absolute
value of the increment in Eq. \ref{Sq}). The next step is to calculate the
scaling of the structure functions, which is defined as the slope, $\zeta(q)$,
measured inside some range of scales where the $S_q(r)$-function is linear and
the field is intermittent. The function $\zeta(q)$ is shown in the upper
right panel in Figure \ref{fig1}. 

%#####################################################################
%#####################################################################

A weak point in the above technique is the determination of the range, $\Delta
r$, where the slopes of the structure functions are to be calculated. To
visualize the range of intermittency, $\Delta r$, we suggest to use the flatness
function (Abramenko 2005b), which is determined as the  ratio of the fourth
statistical moment to the square of the second statistical moment. To better
identify the effect of intermittency, we followed Frisch (1995) and reinforced
the definition of the flatness function and calculated the hyper-flatness
function, namely, the ratio of the sixth moment to the cube of the second
moment:
\begin{equation} 
F(r)=S_6(r)/(S_2(r))^3 \sim r^{ - \kappa}.
\label{Fr}
\end{equation} 
For simplicity, we will refer  to $F(r)$ as the flatness function, or
intermittency spectrum. In case of a non-intermittent structure, the flatness
function is not dependent on the scale, $r$. On the contrary, for an
intermittent structure, the flatness grows as power-law, when the scale
decreases. The slope of flatness function, $\kappa$, determined within $\Delta
r$, i.e., the flatness function exponent, characterizes the degree of
intermittency.

The index  $\kappa$ was computed from the best linear fit within $\Delta r$ in
the double logarithmic plot via the IDL/LINFIT routine by minimizing the
$\chi$-square error statistic. This procedure was repeated for varying
boundaries of $\Delta r$, and a range was determined where the best fit to the
data points has the minimal standard deviation (a range of best linearity). Thus
found range was gradually extended until the change in $\kappa$ and standard
deviation exceed 5\%. This final interval was adopted as $\Delta r$.

\section {Intermittency spectra for various ARs}

\begin{figure}[!h] \centerline { 
\epsfxsize=3.7truein\epsffile{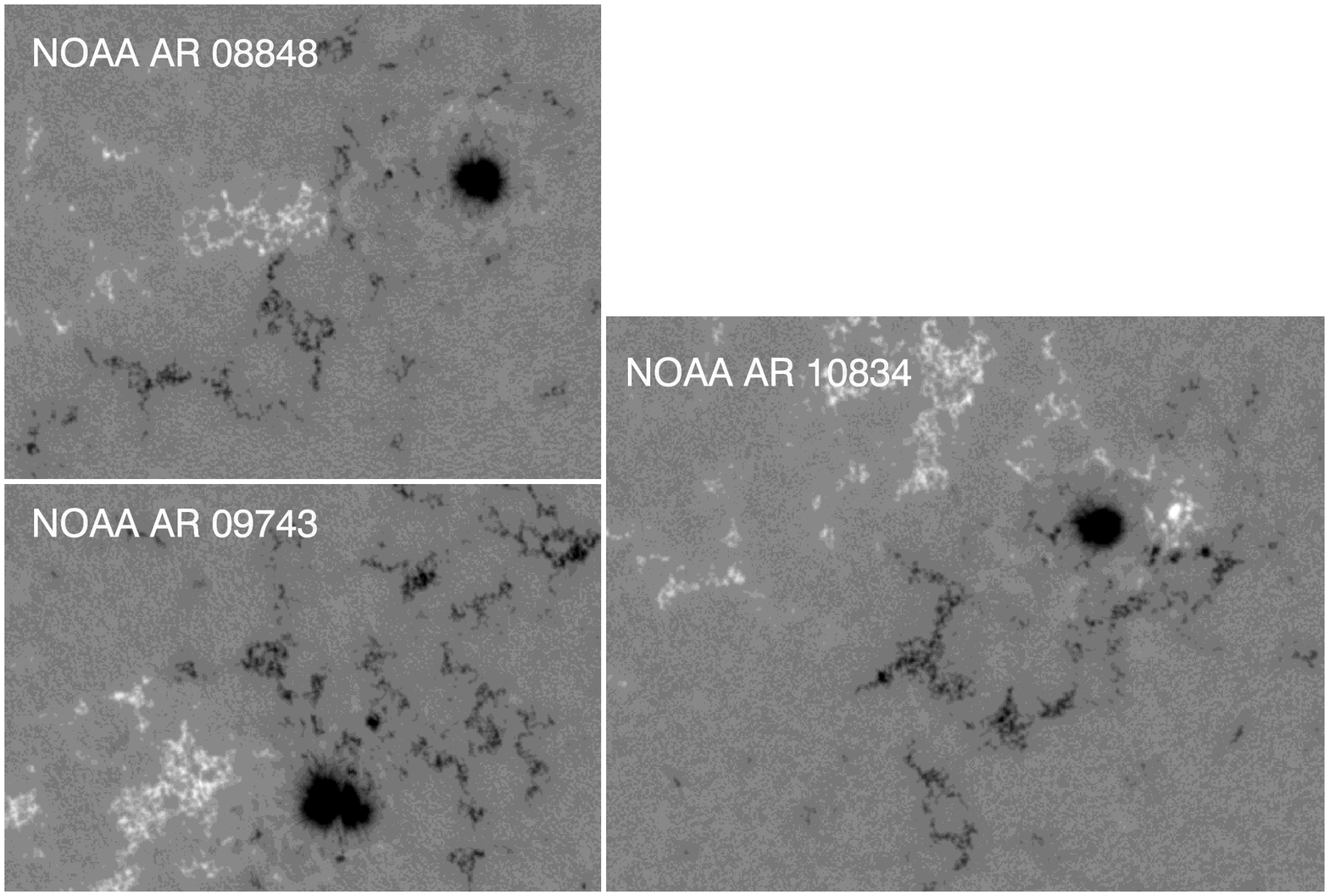}
\epsfxsize=3.3truein \epsffile{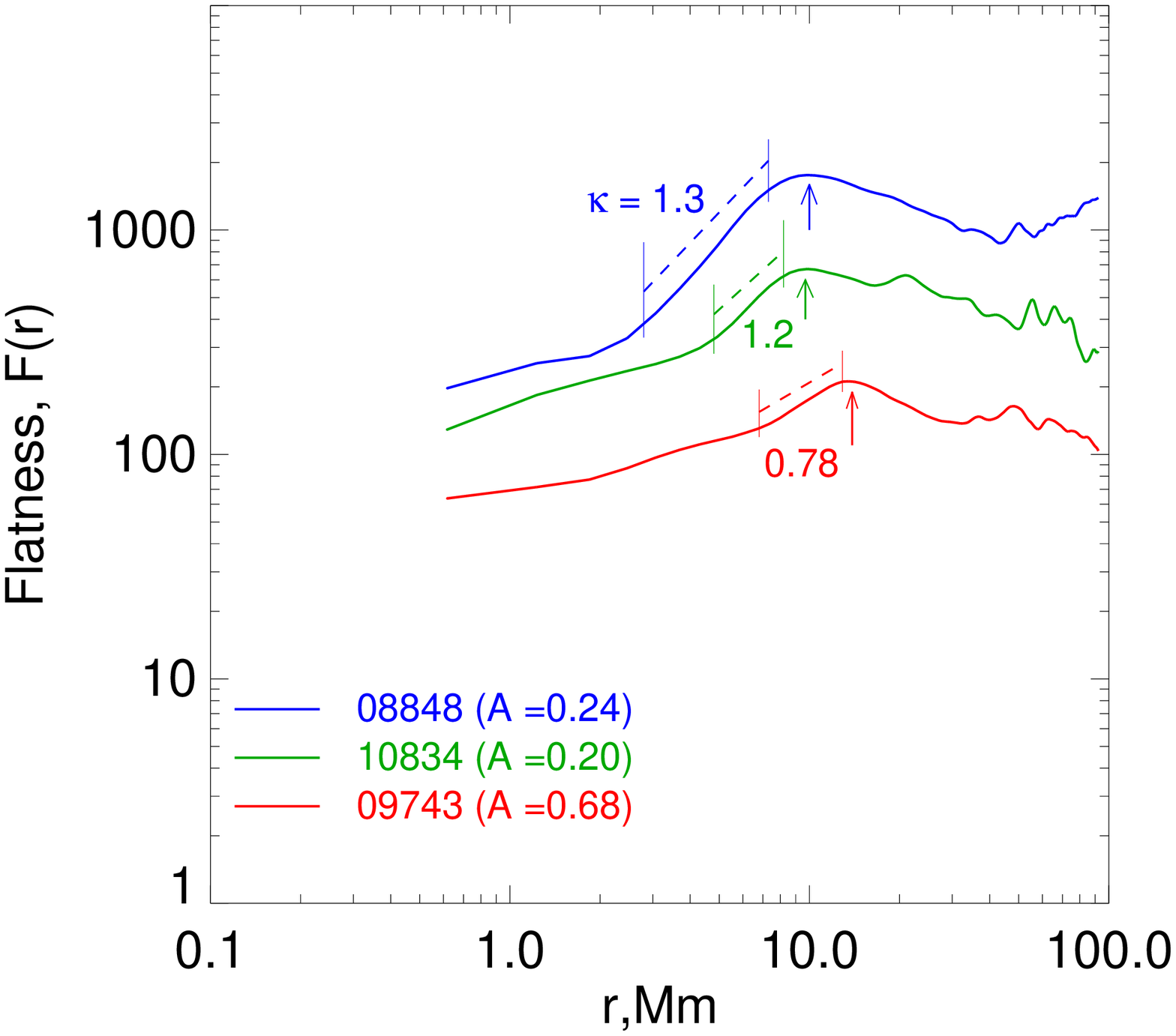}} 
\caption{\sf MDI/HR magnetograms for
three unipolar sunspots groups ({\it left}) and the corresponding flatness
functions ({\it right}). All magnetograms have the same spatial scale, and the
width of NOAA AR 09743 magnetogram is 180 Mm. Flare index, $A$, is
shown for each AR. All flatness functions display a well pronounced maximum at
scales $r_{max}$ (indicated with arrows), roughly corresponding to the size of a
unipolar sunspot. The flatness exponent, $\kappa$, was determined from data
points within the dashed line segments.} 
\label{fig2} 
\end{figure}

\begin{figure}[!h] \centerline {\epsfxsize=3.0truein
\epsffile{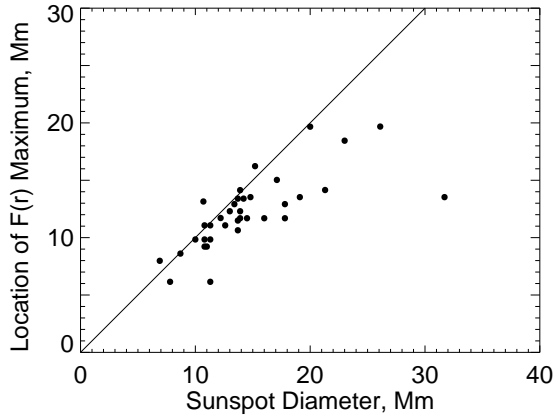}}
\caption{\sf The spatial scale of the maximum, $r_{max}$, in the flatness
function plotted versus the diameter, $d$, of unipolar sunspots. The solid line
is the bisector.}
\label{fig3} 
\end{figure}

Analysis of $F(r)$ functions for various ARs reveals that each AR exhibits a
unique spectrum. The best way to sort them out was to start with simplest and
well defined magnetic structures, such as unipolar spots. Our data set contained
38 unipolar sunspots of different size and polarity. Figure \ref{fig2} shows
magnetograms and flatness functions for three typical unipolar sunspots. The
well pronounced smooth maximum in the flatness function near $r_{max} \approx
10$ Mm is present for all 38 cases.
%===============================================================
 As is evident from Figure \ref{fig3}, the scale, where the maximum is localed,
is well correlated with the size of a unipolar sunspot, $d$. The data points are
predominantly located below the bisector of the diagram, so that the sunspot
diameter is slightly larger than the scale $r_{max}$. The most plausible reason
for that is the saturation effect inside strong sunspots in the MDI
magnetograms, which artificially reduces the size of the latgest magnetic
structure and affects calculations of the structure functions at the scale of
the sunspot size. On the other hand, measurements of the sunspot diameter are
not affected by the saturation effect. 
%-------------------------------------------------------------
According to the definition of intermittency (a structure where rare strong
fluctuations are intermittent with vast areas of low fluctuations), the presence
of an individual sunspot can produce the detectable effect of intermittency at
scales of the sunspot size. Indeed, when we excluded the sunspot area from
calculations of the structure functions, the maximum disappears. Therefore, the
appearance of the smooth maximum in $F(r)$ of unipolar sunspots  is caused by
intermittency introduced by the only strong large sunspot seen in the image.

The positive slope in $F(r)$ at scales smaller than $r_{max}$ (see dashed
segments in right panel of Figure \ref{fig2}) indicates the transition from the
intermittent regime at scales above $r_{max}$ to the non-intermittent regime,
with nearly flat $F(r)$ at scales $r<<r_{max}$. It is natural that the
non-intermittent regime cannot set in abruptly, so that some transition region
appears (as it always does in spectral functions, see, e.g., Frisch 1995). The
most important fact for us here is that there is no {\it negative} slope in
$F(r)$ at $r<r_{max}$, which can be interpreted as an absence of intermittency.
A plateau at small-scales is lower for lower degree of intermittency. Therefore,
the value of the {\it positive} slope of $F(r)$ in the transition region is in
opposite proportion to the degree of intermittency at small scales.

\begin{figure}[!h] \centerline {
\epsfxsize=2.2truein \epsffile{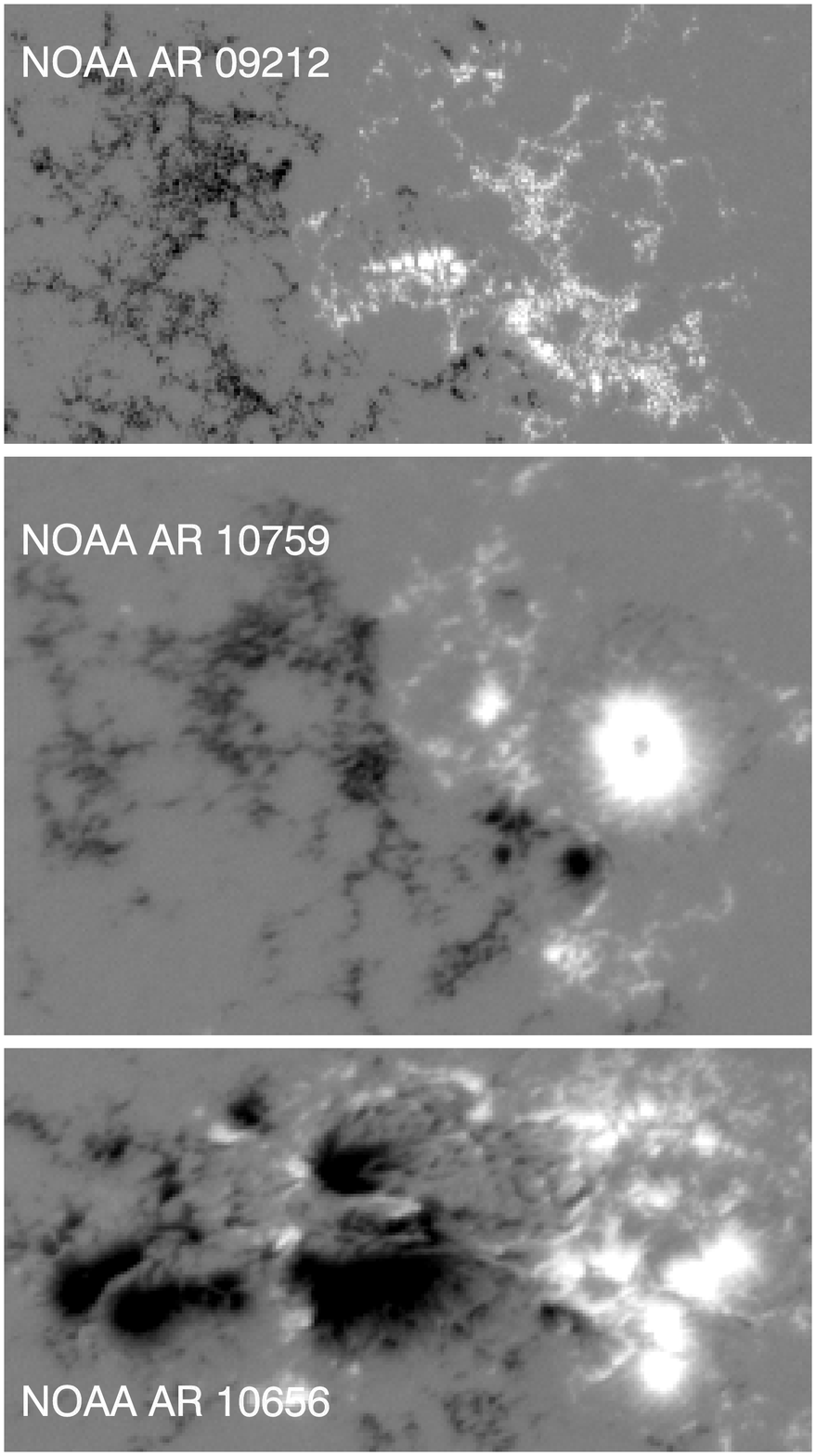}
\epsfxsize=2.6truein \epsffile{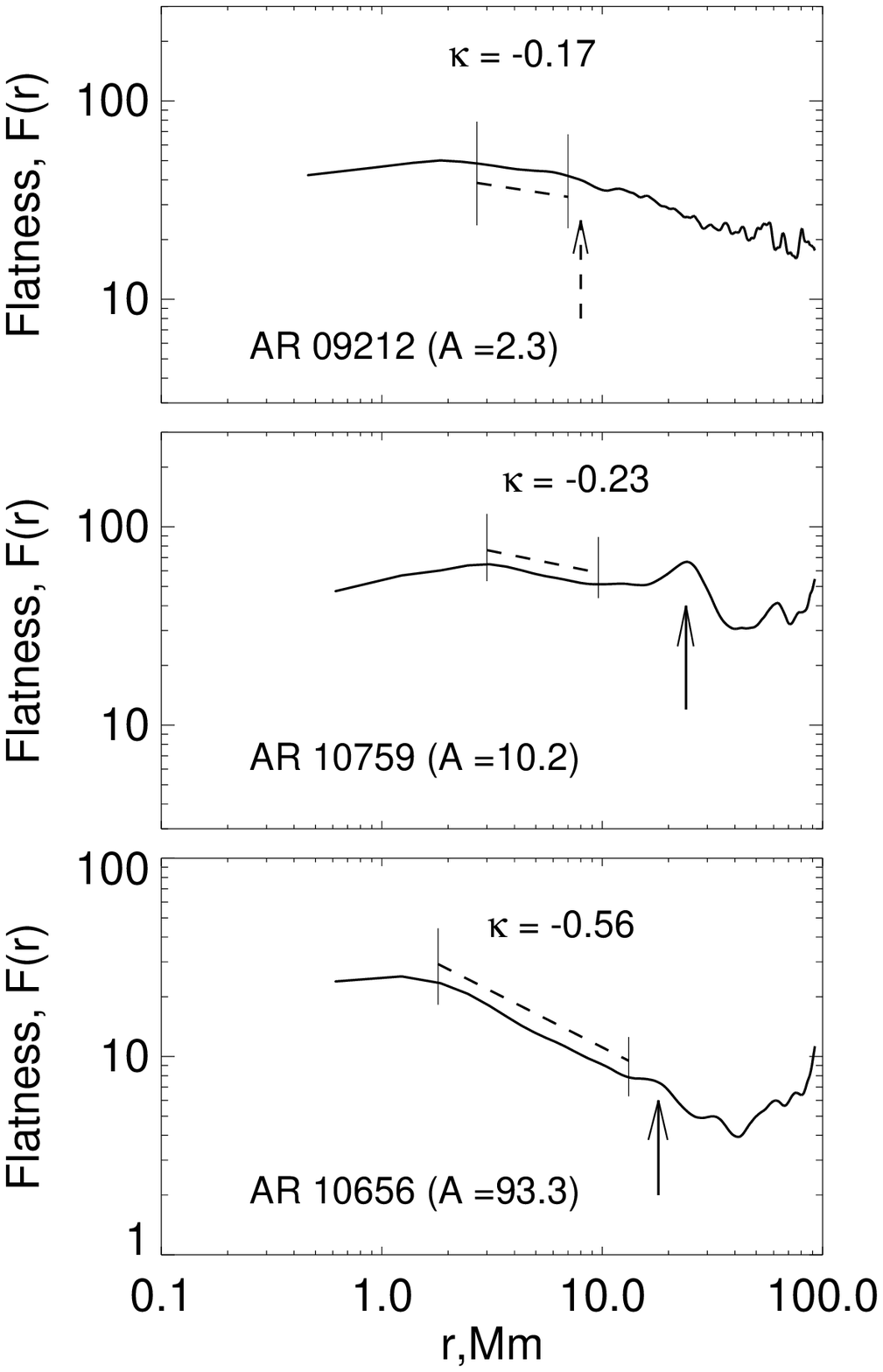}}
\caption{\sf MDI/HR magnetograms for three active
regions ({\it left }) and the corresponding flatness functions ({\it right}). 
The widths of the magnetograms (from top to bottom) correspond
to 290, 170, 183 Mm. The dashed arrow marks a locations of $r_{max}$ derived
as a size of the largest sunspot. Other notations are the
same as in Figure \ref{fig2}.}
\label{fig4} 
\end{figure}

When analyzing more complex ARs, one might expect the presence of a local
maximum in $F(r)$ at scales of about 10-20 Mm, which appears to be a
manifestation of large-scale intermittency introduced by strong isolated
sunspots. We determined the diameter, $d$, of the largest sunspot (as an average
half-width of $B_l$ crosscut made in two orthogonal directions across the
sunspot center) in each AR and then found the closest local maximum in $F(r)$
and adopted it as $r_{max}$ (see Figure \ref{fig4}, middle and bottom rows).
When there was no local maximum, we accepted $r_{max}=d$ (Figure \ref{fig4}, top
row). We consider the position of $r_{max}$ as a boundary between the
large-scale intermittency (related to the presence of strongest sunspots) and
small-scale intermittency (related to the internal magnetic complexity in an
AR), and we use it below.

%#####################################################################
%#####################################################################
%#####################################################################
%#####################################################################

%#####################################################################
%#####################################################################

%#####################################################################
\begin{figure}[!h] \centerline {
\epsfxsize=3.5truein \epsffile{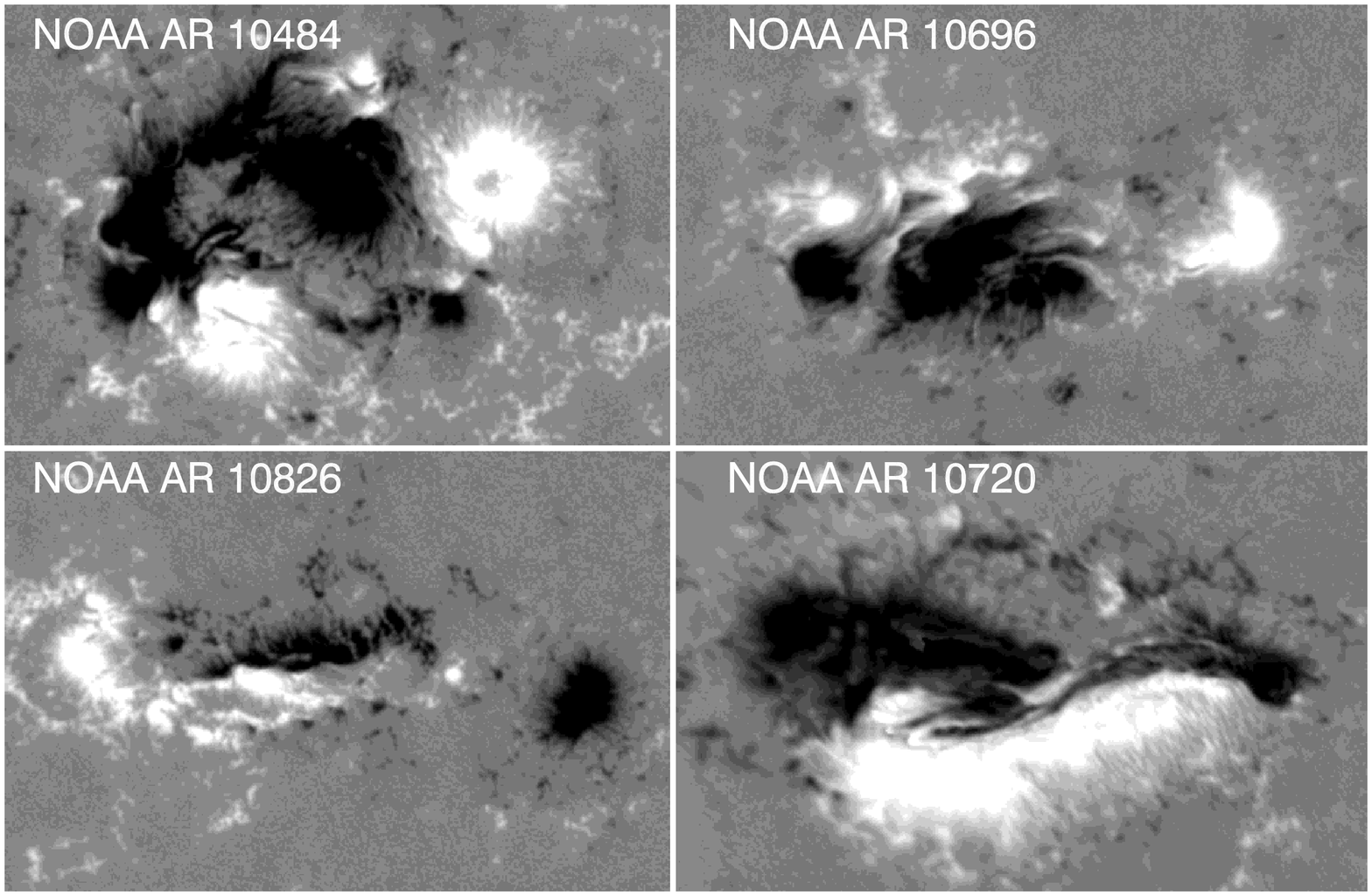}
\epsfxsize=3.0truein \epsffile{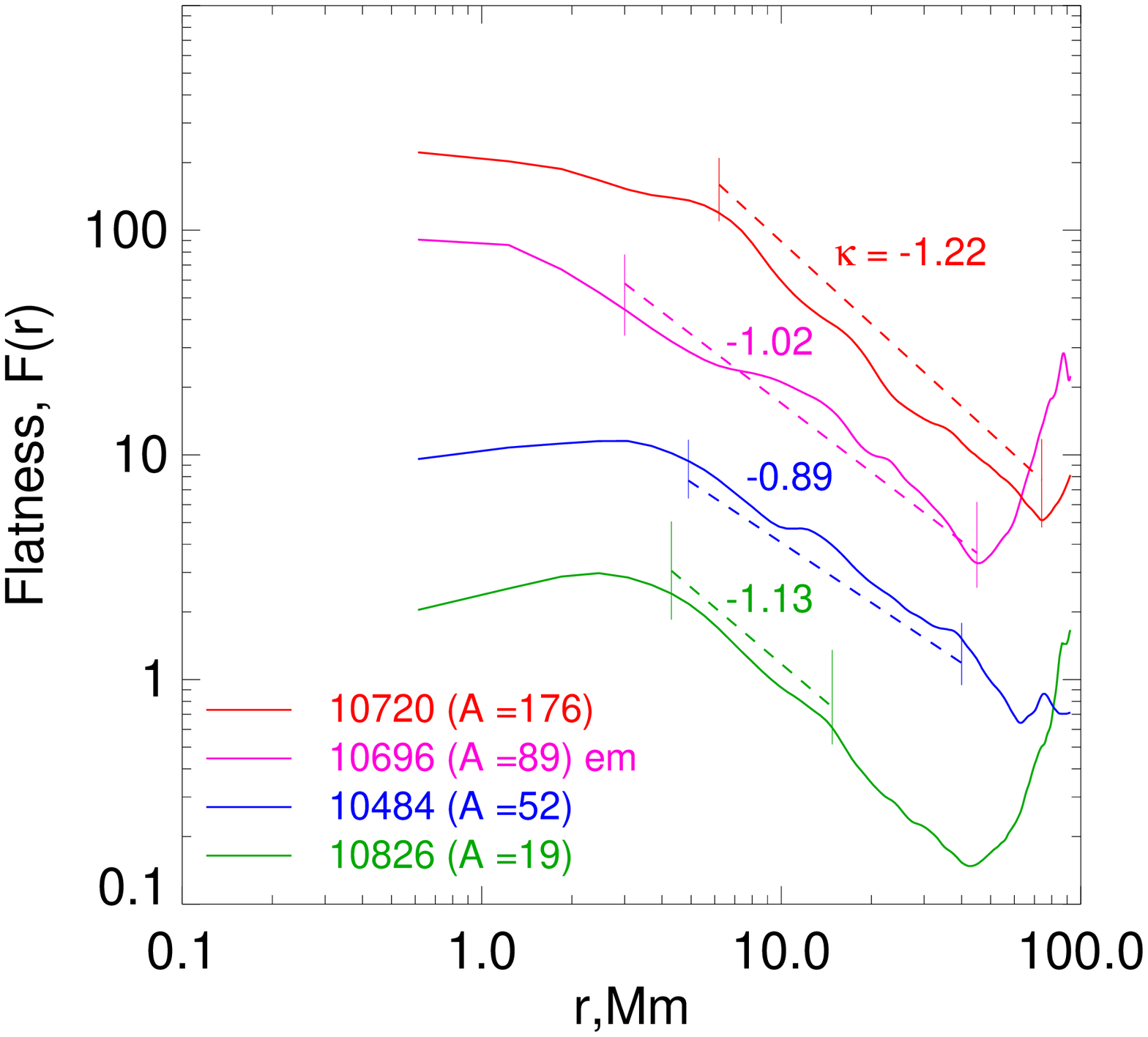}}
\caption{\sf MDI/HR magnetograms ({\it left}) and flatness functions ({\it
right}) for four ARs with the most pronounced intermittency at small scales.
$F(r)$ functions are shifted along the $Y$-axis into "well-readable" positions.
The dashed line segments mark ranges of small-scale intermittency. Flare index,
$A$, and exponents $\kappa$ are indicated for each AR. All magnetograms have the
same spatial scale, and the width of the AR NOAA 10484 magnetogram is 165 Mm.}
\label{fig5} 
\end{figure}
%#####################################################################

Flatness functions for ARs, which are more complex than unipolar sunspots,
showed an undulating behavior and presence of negative-slope regions, $\Delta
r$, at scales below $r_{max}$ (see Figure \ref{fig4} for typical examples). In
this case, intermittency at scales within the range $\Delta r$ is not defined by
the presence of large sunspots but is rather caused by smaller-scale complexity
associated with small sunspots and $\delta$-structures, strong gradients along
numerous neutral lines, channel-like structures in the magnetic field, etc.
Also, we note a tendency that in more complex ARs the intermittency interval
and the steepness of the spectrum become larger.

We found 12 ARs, for which the flatness function was extremely steep and closely
followed power law at all scales below 40-70 Mm down to 2-5 Mm. Examples of
extremely intermittent cases are shown in Figure \ref{fig5}, where very broad
ranges of steep power-laws in $F(r)$ are evident, which is typical for highly
intermittent structures (Frisch 1995). These ARs are the most complex in our
data set and frequently they represent large $\delta$-structure active regions.

\section {Intermittency Index versus Flare Index }

\begin{figure}[!h] \centerline {
\epsfxsize=3.4truein \epsffile{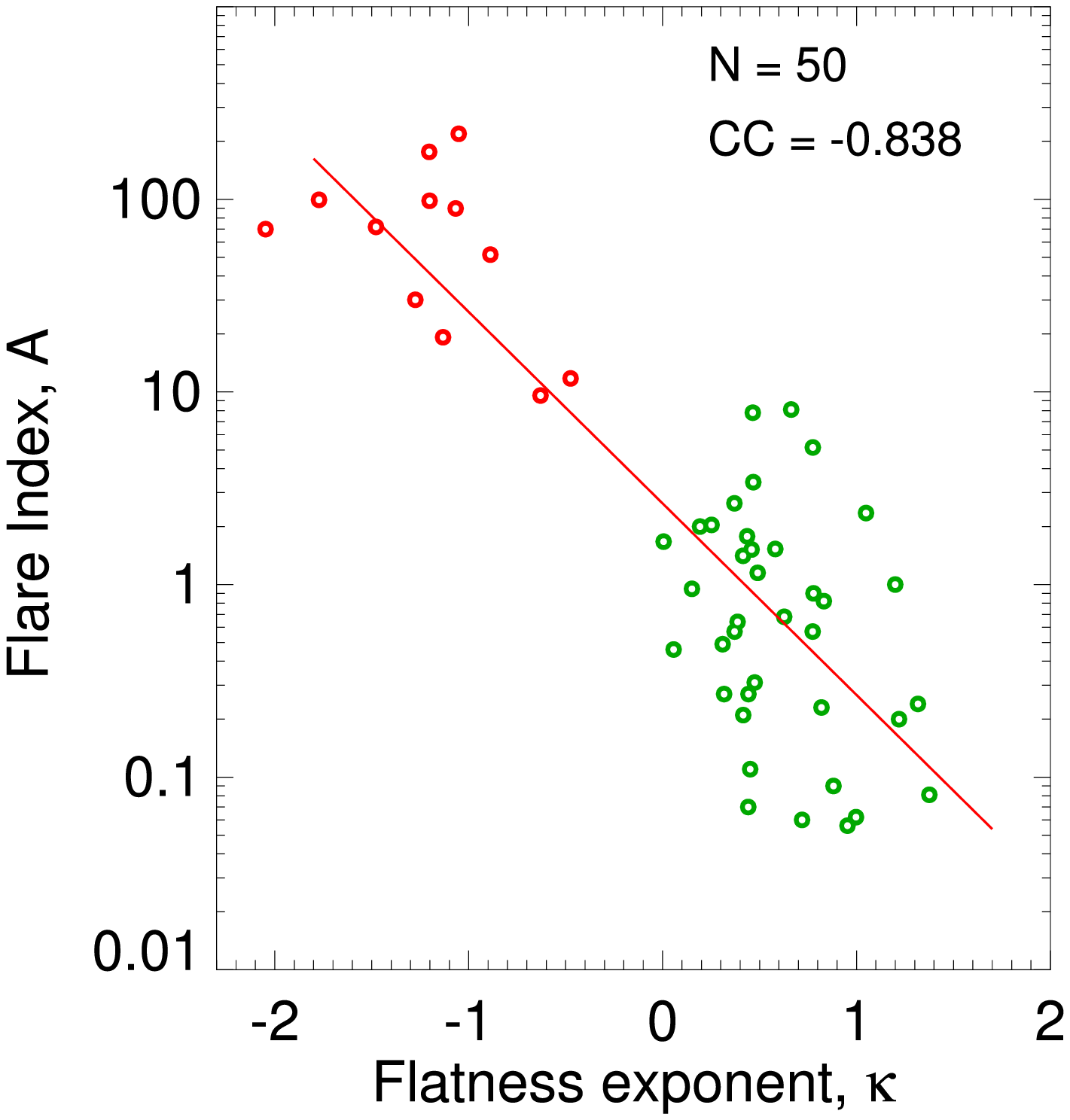}
\epsfxsize=2.9truein \epsffile{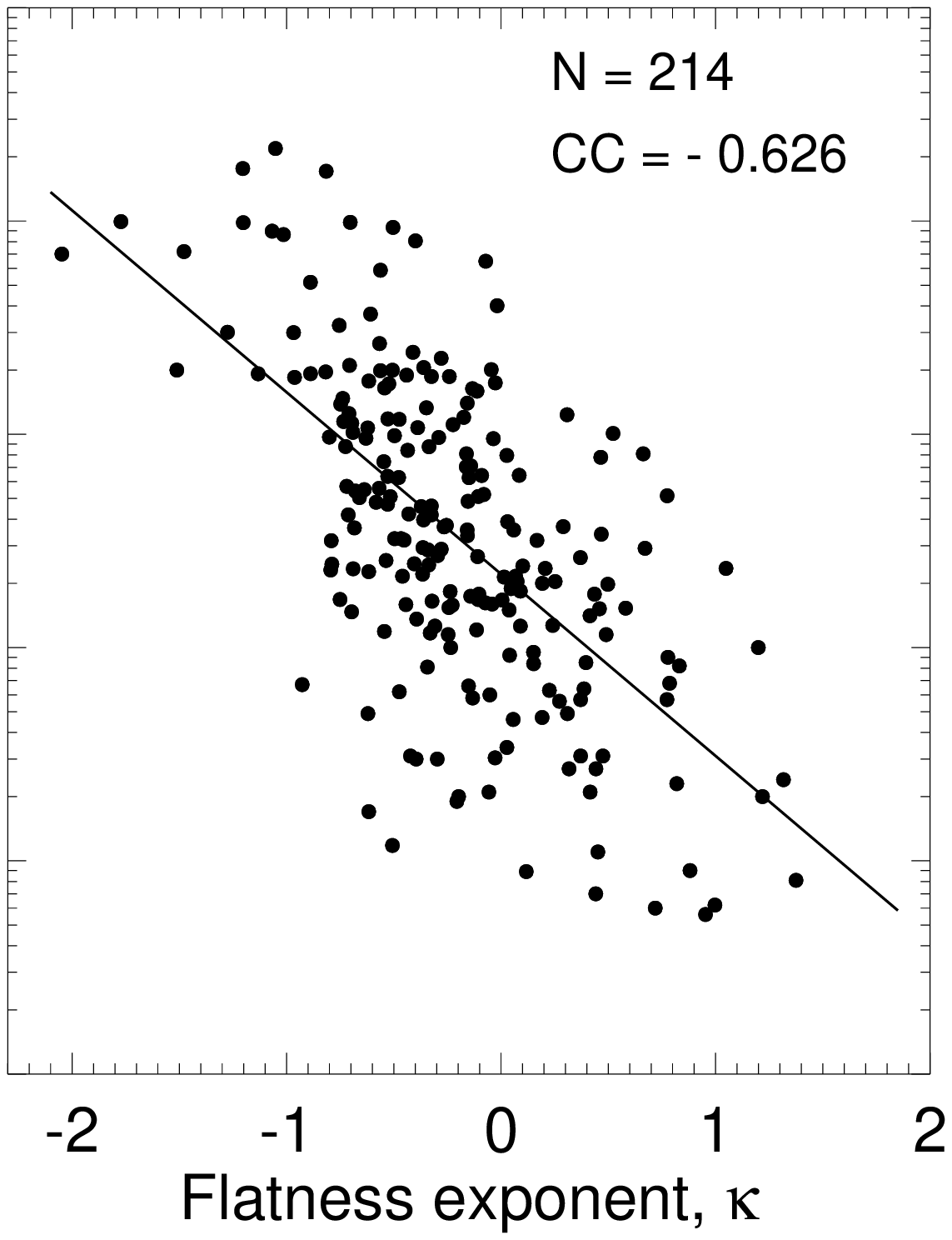}}
\caption{\sf {\it Left -} The flare index, $A$, plotted versus the flatness
exponent, $\kappa$, for 12 ARs of the most steep and extended intermittency
spectra (red circles) and for 38 non-intermittent ARs (green circles).
{\it Right -} The same but for all ARs in the data set.}
\label{fig6} 
\end{figure}

In Figure \ref{fig6} ({\it left}) we related the flatness exponent, $\kappa$,
with the flare index for 12 ARs of highest intermittency and for 38
non-intermittent ARs  - unipolar sunspots. The non-intermittent ARs (green
circles) display very low flare productivity, whereas the ARs of highest
intermittency (red circles) all are associated with very high flare index. the
Pearson correlation coefficient is $-0.83$ (with the 95\% confidence interval of
0.79 - 0.87 according to the Fisher's Z-transformation statistical test of
significance\footnote{http://icp.giss.nasa.gov/education/statistics/page3.html}
). Thus, the non-intermittent and highly-intermittent ARs are prone to display
low and high flare productivity, respectively.

The right panel in Figure \ref{fig6} is a combined plot for all ARs. The
tendency here is the same: higher degree of intermittency corresponds to
stronger flare productivity. For all ARs, the Pearson correlation
coefficient is $-0.63$ with the 95\% confidence interval of 0.54 - 0.70.

%============================================
Apart from our data set of 214 flaring ARs, there was a subset of 34 ARs
(also observed during 1997-2006) for which the total flux exceeded the threshold
value of 10$^{22}$ Mx while the respective flare index was zero. Most of them
were unipolar sunspot groups. We find that their flatness function exponent,
$\kappa$, is in the range of 0 - 1.5 which is practically the same as that for
low-flaring ARs. Future observations with better flare and flux sensitivity
might change the right-hand part of the diagram in Figure \ref{fig2}{\it b}, but
not much changes are expected for the top-left part of the diagram. 

%--------------------------------------------------

We also compared the flare index versus the slope of the spectrum at scales {\it
exceeding} $r_{max}$ and obtained a very low correlation (CC=-0.25
with the 95\% confidence interval of 0.12 - 0.37). Therefore, the intermittency
at large scales hardly contributes to the flare productivity of an AR. It rather
indicates the presence of dominant sunspots in the AR. To the contrary,
small-scale intermittency (at scales $r < r_{max}$) seems to be essentially
related to the AR's flaring rate. Further we will discuss intermittency only at
small scales.

%#####################################################################
%#####################################################################

\section {Flatness Functions from MDI/HR and Hinode SOT/SP Magnetograms}

\begin{figure}[!h] \centerline {\epsfxsize=4.0truein
\epsffile{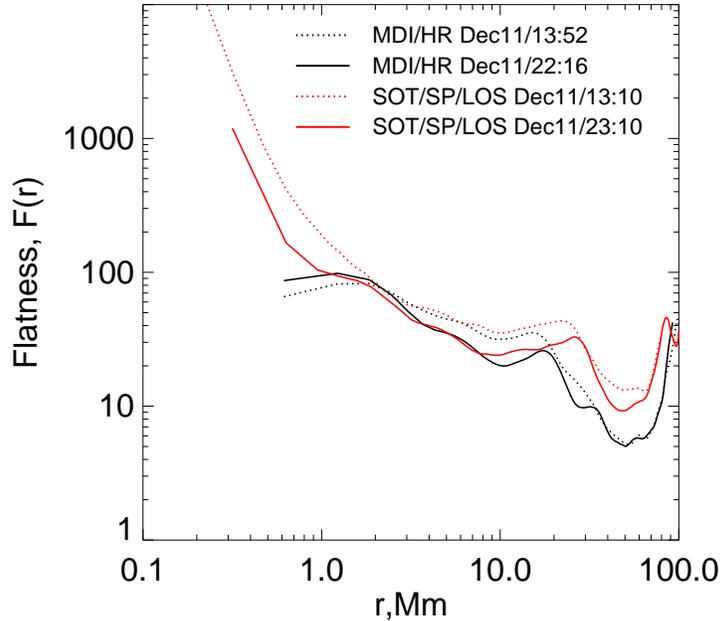}}
\caption{\sf Flatness functions calculated from two MDI/HR magnetograms ({\it
black}) and two Hinode SOT/SP LOS magnetograms ({\it red}) for NOAA AR 10930. 
Dotted (solid) lines refer to the magnetograms taken approximately 39 (27) hours
prior to the X3.4 flare. The 13:10 SOT/SP magnetogram  ({\it red dotted}) has a
pixel size of 0.148$ \times$ 0.16 $^{''}$ and shows the steepest growth of
$F(r)$ at small scales. The 23:10 SOT/SP magnetogram ({\it red solid}) has a
pixel size of 0.297$ \times$ 0.32 $^{''}$. }
\label{fig7} 
\end{figure}

We compared flatness functions for NOAA AR 10930 derived with SOHO/MDI and
Hinode/SOT/SP magnetographs observed on December 11, 2006 (Figure \ref{fig7}).
The level2 SP magnetograms were processed with HAO's MELANIE code. The data are
available at $ http://sot.lmsal.com/data/sot/level2hao-new $. At scales 2 - 10
Mm, the two functions are in good agreement. At scales below 2 Mm, the MDI
flatness functions saturate, while Hinode flatness functions show rapid increase
with decreasing scale. The slope there is even steeper than it is inside the  2
- 10 Mm range, which indicates that intermittency might become stronger as the
scale decreases and the spatial resolution improves. Note that the increase
of the Hinode resolution results in the steepening of the flatness function
at small scales. This comparison shows that
the saturation in the MDI functions is caused by low sensitivity and resolution
of an instrument rather than by absence of intermittency at small scales. 

%#####################################################################
%#####################################################################

\section { Directional Structure Functions }

As we indicated earlier, strong field gradients along a neutral line, small
$\delta$-structures and shredded magnetic fields all can be plausible cause of
strong small-scale intermittency. In attempt to clarify this assumption, we
refined the calculation of the structure functions as described below.

According to the general definition of structure functions (see Section 3),
functions $S_q(r)$ produce an intermittency measure, which is averaged over all
possible directions of the separation vector ${\bf r}$. To explore how the
intermittency varies in different directions, we calculate directional structure
functions, where the averaging in Eq. \ref{Sq} is performed only over the
separation vectors of a certain orientation, ${\bf r}(\theta)$ (see lower right
panel in Figure \ref{fig1}, the origin of ${\bf r}$ is arbitrary). Specifying a
set of bins for $\theta$ ranging from
0$^\circ$ to 180$^\circ$ we calculated $S_q(r,\theta)$ functions.
%#####################################################################
\begin{figure}[!h] \centerline {
\epsfxsize=1.8truein \epsffile{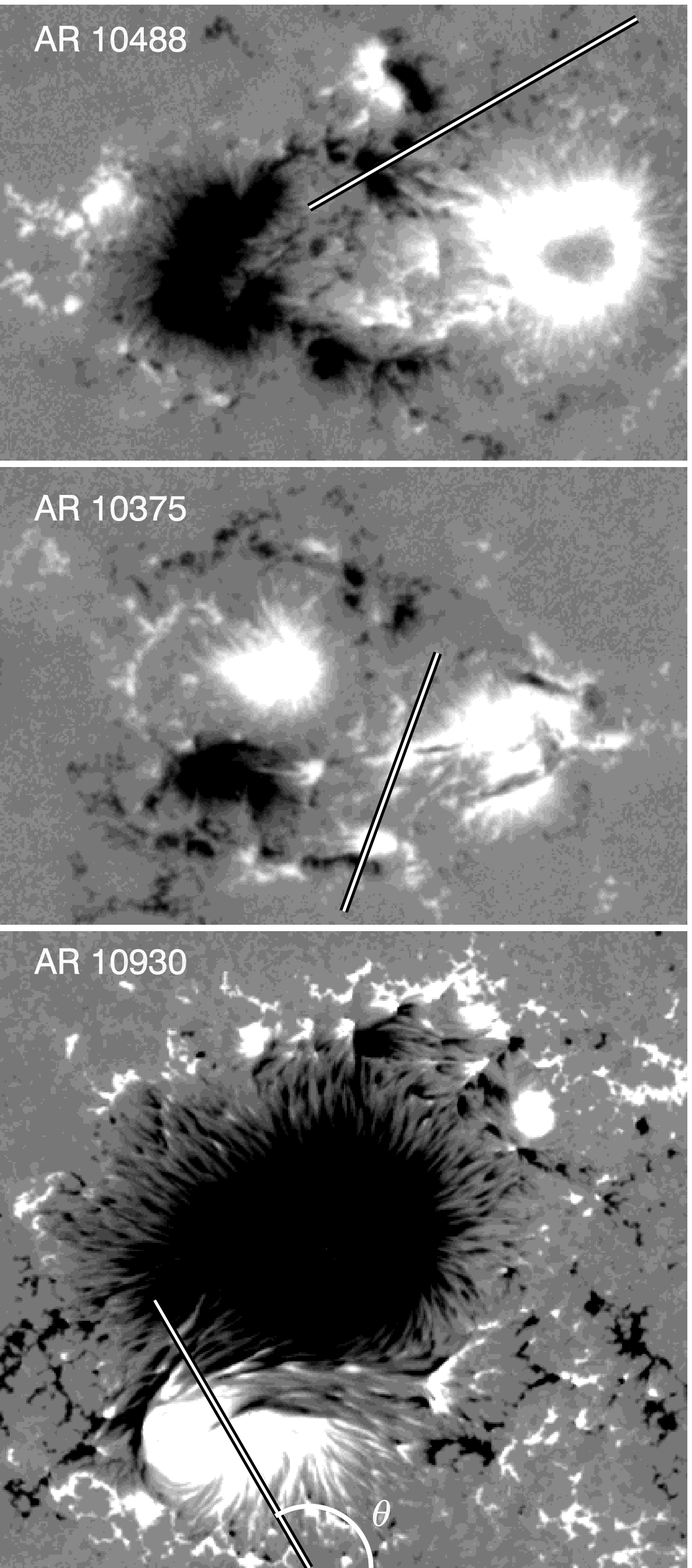}
\epsfxsize=3.6truein \epsffile{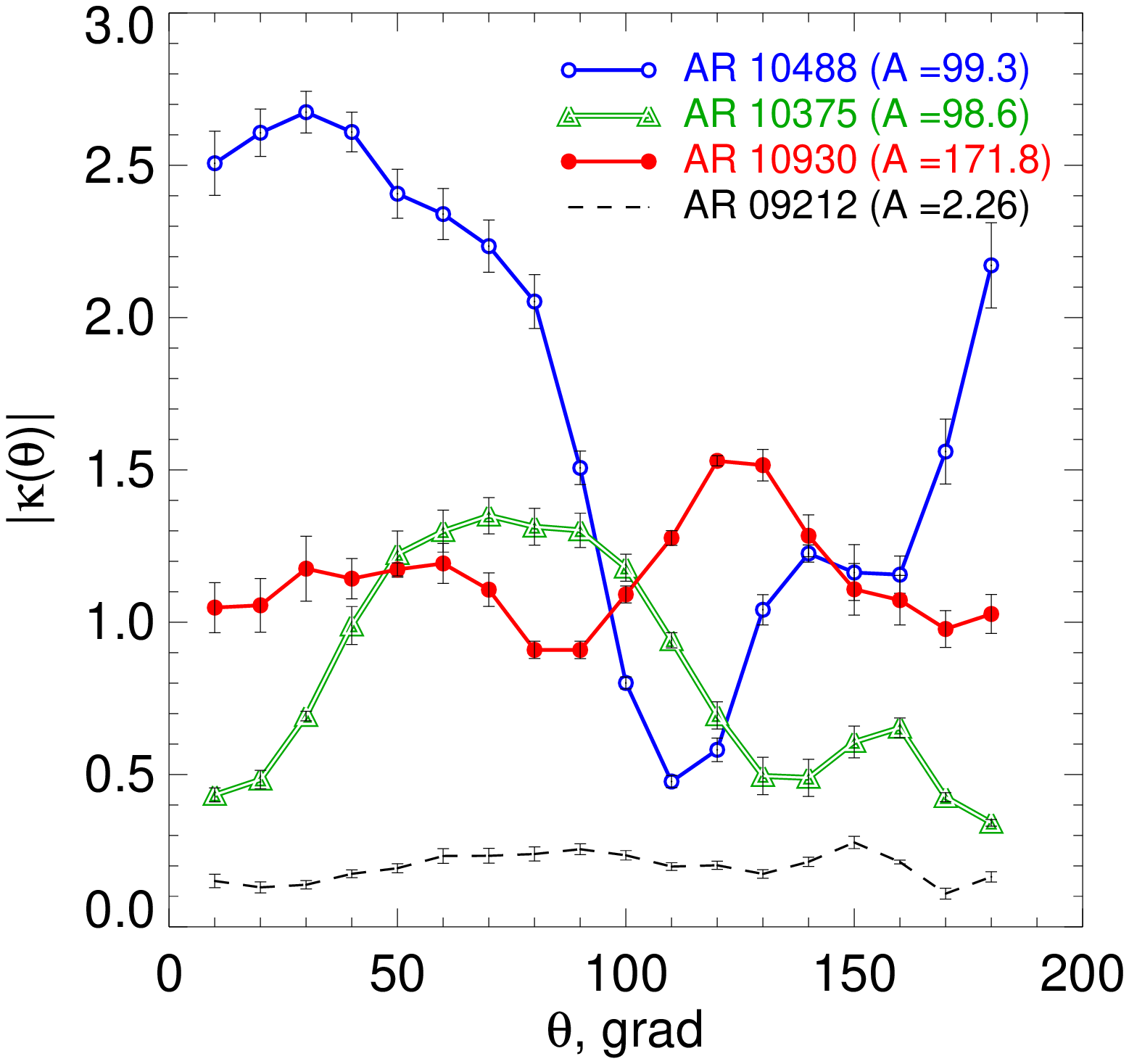}}
\caption{\sf {\it Left - } MDI/HR magnetograms of NOAA ARs 10488 and 10375 (two
top panels) and SOT/SP LOS magnetogram of NOAA AR 10930. All magnetograms are
scaled between -800 and 800 G. North is to the top, west is to the right. The
width of the MDI magnetograms is 152 Mm and SP magnetogram is 77 Mm.
{\it Right - } Variations of the modulus of $\kappa$ versus the directional
angle, ${\theta}$, measured in the counter-clock-wise direction from the EW
direction of the horizontal axis on the corresponding magnetogram. The
orientation $\theta(max)$, where each function $|\kappa(\theta)|$ peaks, is
indicated with double line segments in the magnetograms. For the best view, the
origins of segments are chosen so that the segments intersect the features of
interest.}
\label{fig8} 
\end{figure}
%#####################################################################
We thus find that $|\kappa(\theta)|$ varies significantly with $\theta$.
Moreover, these fluctuations are particularly strong in the ARs showing strong
flare productivity. Examples are shown in Figure \ref{fig8}. Thus, for NOAA AR
10488, the angle segment bounded by directions of $0^\circ$ and $70^\circ$  with
the maximum at $\theta(max)=30^\circ$ is associated with much stronger
intermittency as compared to other directions. When we plot this preferred
direction ($\theta(max)=30^\circ$) on the magnetogram (double line segment in
the upper left panel of Figure \ref{fig8}), we find that it crosses a set of
small-scale extended threads and filament-like magnetic features connecting the
two opposite polarity sunspots. Besides, it is also the direction of the largest
gradient across the small delta-structure in the northern part of the AR.
Similar situation also appears in several other high-flaring ARs (Figure
\ref{fig8}). We found more symmetric angular distribution of intermittency in
ARs of low flare productivity. An example for AR NOAA 09212 is shown with a
dashed line in Figure \ref{fig8} ({\it right}). 

\section {Connection to the multifractality spectrum, $f(\alpha)$ }

The proposed technique to calculate flatness function, $F(r)$, also allows us to
determine the intermittency interval, $\Delta r$. This interval appears to be
the best appropriate interval to derive the slope, $\zeta(q)$, of the $S_q(r)$
function. Indeed, scaling of the structure function should be determined at an
interval where the field is intermittent. However, in very smooth $S_q(r)$
functions, there is no a priory indications for such an interval (see Figure
\ref{fig1}). It can, fortunately, be determined by using function $F(r)$: the
intermittency interval is detected as a range where $F(r)$ increases as a power
law with decreasing scale. The slope of $S_q(r)$, derived inside $\Delta r$ for
each $q$, gives us $\zeta(q)$ function (Figure \ref{fig1}, upper right panel).

For monofractals, function $\zeta(q)$ is a straight line due to a global
scale-invariance. And it has a concave shape in case of a multifractal. The
degree of concavity is usually measured by function $h(q) =\zeta(q) /dq$. All
values of $h$, within some range, are permitted for a multifractal, and for each
value of $h$ there is a monofractal with an $h$-dependent dimension $D(h)$ at
which the scaling holds with exponent $h$ called the strength of singularity.
This representation of multifractality is based on the increments of the field
and has its roots in the theory of turbulence (Kolmogorov 1941). There exists
another representation based on the dissipation, $\varepsilon$, of the field
energy, which relies on the Kolmogorov's (1941) result stating that field
increments over a distance $r$ scale as $(\varepsilon r)^{1/3}$, known as the
refined similarity hypothesis (Monin\& Yaglom 1975). In multifractal terminology
the refined scaling hypothesis means that to any singularity of exponent
$\alpha$ of $\varepsilon r$, there is an associated singularity of exponent
$h=\alpha/3$ for the field of the same set, which has the same dimension $D(h)$.
Usually, it is very difficult to measure the local dissipation in the 3D space.
So, one-dimensional space averages of the dissipation are usually used. The
corresponding dimension $f(\alpha) = D(h) - (d-1)$ has decreased by two units
(for the space dimension $d=3$) because one-dimensional cuts of a 3D structure
are taken. In the literature $f(\alpha)$ is often referred as the
multifractality spectrum (see, e.g., Feder 1988; Lawrence et al. 1993; Frisch
1995; Schroeder 2000; Conlon et al. 2008; McAteer et al. 2009).
The values of $D(h)$, in turn, can be  calculated as a Legendre
transform of $\zeta(q)$ (Frisch 1995):
\begin{equation}
D(h(q)) = inf_{q}(d + q h(q) - \zeta(q)). 
\label{Dh} 
\end{equation}  
When $\zeta(q)$ is concave, then for a given
value of $q$ (recall that $q$ takes real values) the extremum in Eq. \ref{Dh} is
attained at the unique value $h_{o}(q)$, and 
\begin{equation}
D(h_{o}(q)) = d + q h_{o}(q) - \zeta(q)).
\label{Dh2} 
\end{equation}

\begin{figure}[!h] \centerline {
\epsfxsize=3.0truein \epsffile{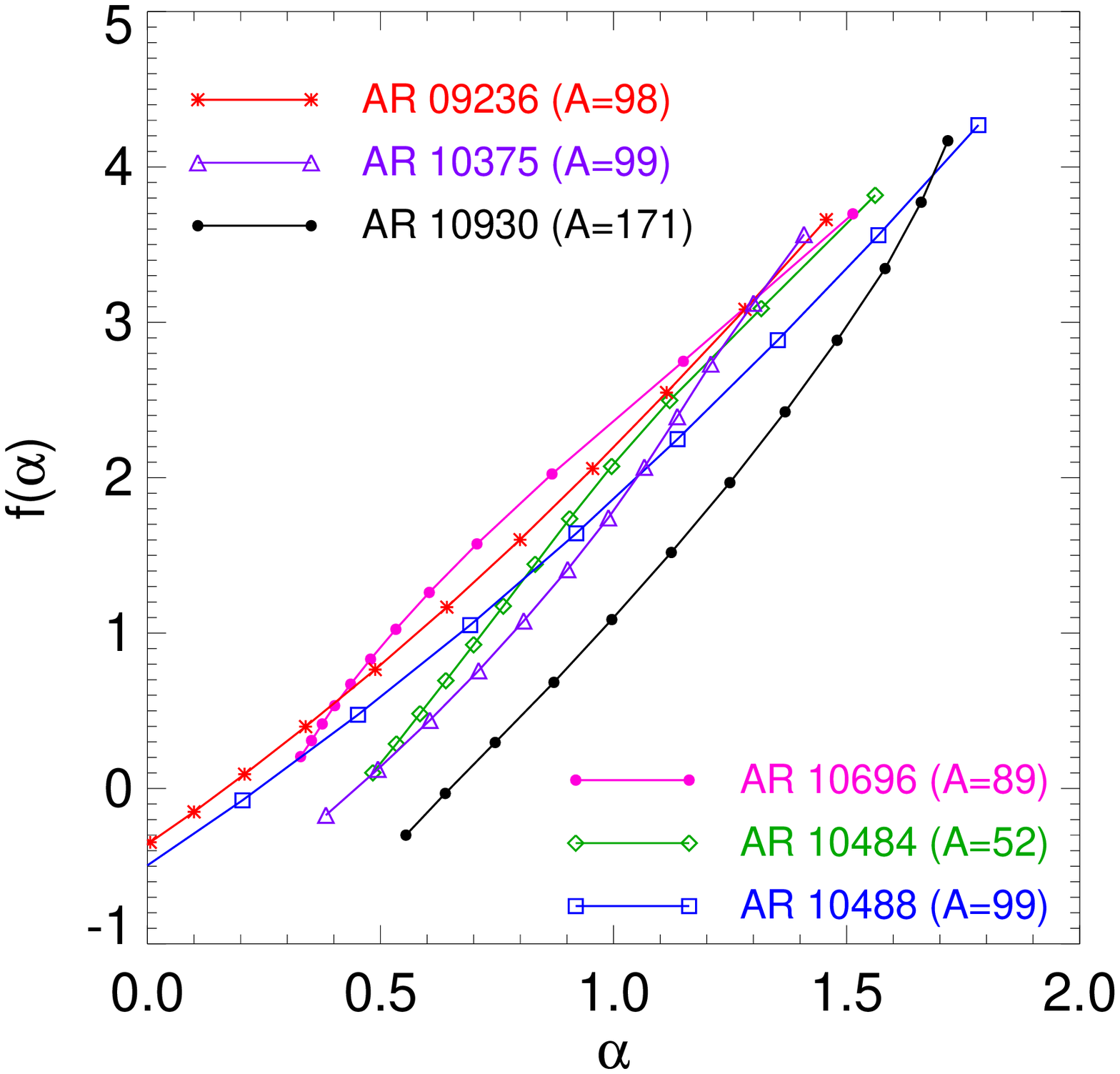}
\epsfxsize=3.0truein \epsffile{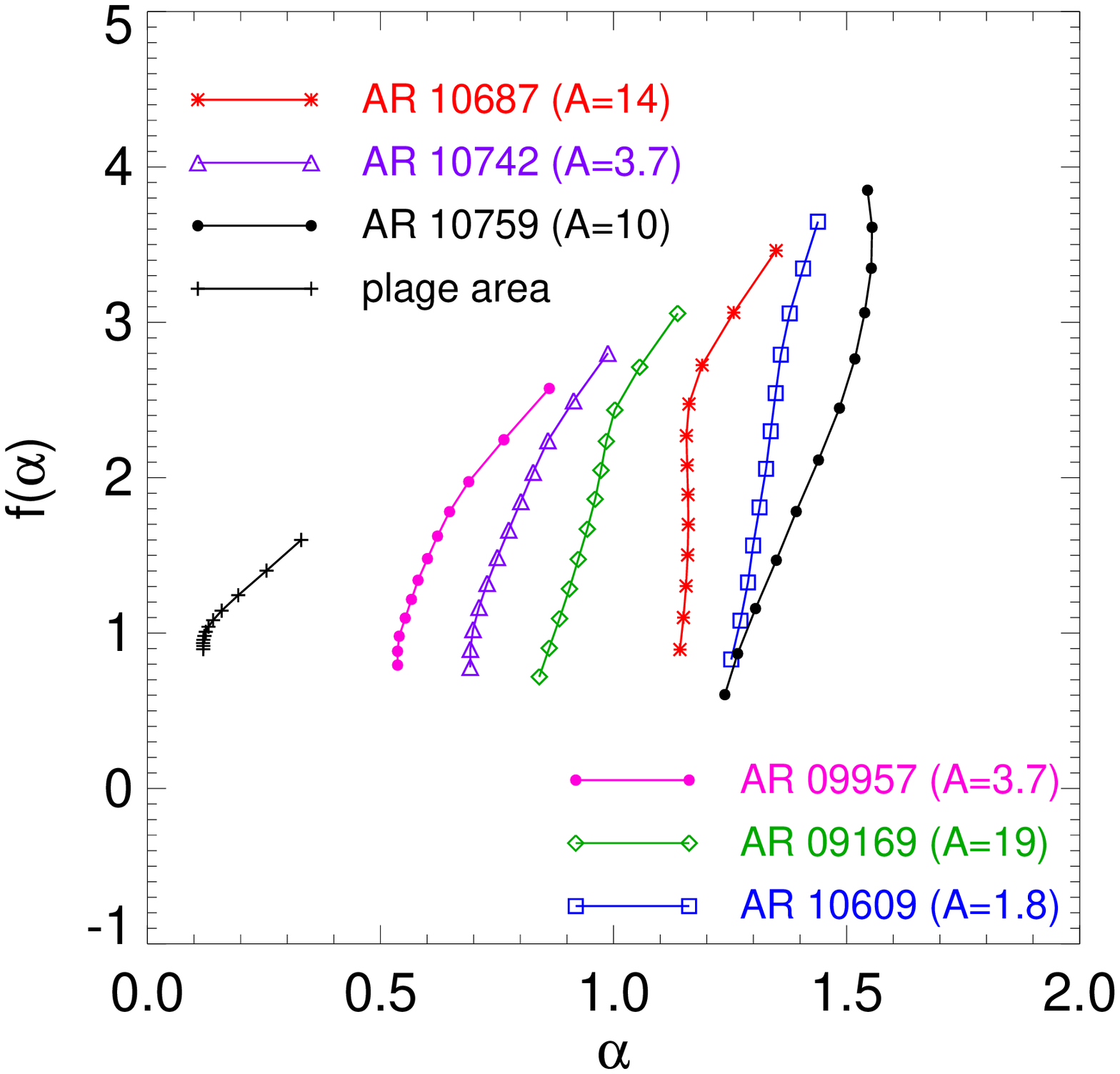}}
\caption{\sf 
Multifractality spectra, $f(\alpha)$, plotted: {\it left - } for ARs of
high flare productivity; {\it right -} for ARs of low flare productivity and
for a plage area. 
Spectra on the left frame are more broad than that on the right frame.}
\label{fig9} 
\end{figure}

We applied the above formulas for our case, $d=2$.
Examples of the multifractality spectra are shown in Figure
\ref{fig9}. One can see that the most complex and flare-productive ARs
(left frame in Figure \ref{fig9}) exhibit broader spectra as compared to
that of low-flaring ARs. This means that a set of monofractals that
form an observed multifractal, is much more broad in highly-flaring ARs as
compared to low-flaring ARs. 
For comparison, in Figure \ref{fig9} (right frame) we plot $f(\alpha)$
calculated for a plage area. Flatness function for plage areas are
linear in a broad range of scales, with small $\kappa$, - similar to that shown
in Figure \ref{fig4} (top right panel). However, the multifractality spectrum
for them is rather narrow, similar to that for low-flaring ARs.

%#####################################################################
%#####################################################################

\section { Conclusions and Discussion}

Analysis of intermittency and multifractality for magnetic structure in 
ARs allows us to conclude the following.

1. Intermittency at small scales is relevant to flare productivity. For the
majority of ARs there is a smooth maximum in the intermittency spectrum at
scales of $r_{max} \approx 10$ Mm. The characteristic scale value $r_{max}$
roughly corresponds to the size of dominant sunspots in an AR. Two intervals of
intermittency on both sides of $r_{max}$ were defined. Namely, large-scale
intermittency at scales above $r_{max}$, and small-scale intermittency at lower
scales. The large-scale intermittency is caused by the presence of strongest
magnetic inhomogeneities - dominant sunspots. The degree of the large-scale
intermittency does not correlate with flare productivity. The small-scale
intermittency is present in all ARs except unipolar sunspots. The small-scale
region of intermittency spectra is steeper for ARs of higher flare
productivity. The correlation coefficient between the flare index and the index
of the small-scale intermittency spectra is - 0.63.

2. In respect to the shape of intermittency spectra, all ARs could be divided
into three types. The ARs of the first type (12 out of 214, most of them are
$\beta\gamma\delta$-type ARs according to Mount Wilson classification)
display a well-pronounced power-law steep spectrum in a broad range of scales,
$\Delta r$: from 40-70 Mm down to 2-5 Mm. For some of them, there is no
characteristic scale $r_{max}$, and the large-scale intermittency interval is
smoothly connected to the small-scale intermittency interval. These ARs are very
complex and the most flare productive. They also display the highest degree of
intermittency as measured by the flatness exponent, $\kappa$, and the broadest
multifractality spectrum, $f(\alpha$), (see below).

The second type of ARs (164 of 214) are moderate active regions (most of them
are $\beta\gamma$-type ARs). Their spectra
usually display the $r_{max}$-peak followed by the moderate increase of $F(r)$
at smaller scales. The flatness exponent in these ARs tends to be lower than
that for the ARs of the previous type.

The third type of ARs (38 out of 214) do not display intermittency at scales
 $r < r_{max}$. All of these ARs are unipolar spots ($\alpha$-type ARs
according to Mount Wilson classification) of very low flaring rate.

3. Analysis of directional intermittency spectra determined for various
directions on a magnetogram showed that in flare productive ARs, angular
variations of intermittency are highly inhomogeneous. There are angular segments
where the degree of intermittency is highest, and these contribute significantly
into the total amount of intermittency in an AR. These directions cut across the
main direction of narrow magnetic filaments and/or along the largest gradients
in a small $\delta$-structures present in an AR. In NOAA AR 10930, for example,
this direction is nearly perpendicular to the magnetic filaments along the main
neutral line separating two sunspots of the global delta-structure. We suggest
that these small-scale shredded and stressed magnetic filaments are responsible
for formation of small-scale intermittency and they are relevant to the high
flare rate. This idea seems to be in good agreements with Schrijver et al.
(2008) who reported that in this particular event, a current-carrying twisted
flux rope was responsible for the powerful X3.4 flare. 

4. The important outcome of the above study is that the most flare-productive
ARs tend to represent the most abundant multifractals. The multifractality
spectra, $f(\alpha)$, show that ARs of highest flare productivity display the
broadest range of multifractality (the range of singularity exponents, $\alpha$
and $h$). This further indicates that their magnetic structure consists of a
voluminous set of monofractals, and this set is much richer than that for
low-flaring ARs. Changes in the multifractality spectra during the AR's
emergence were studied by Conlon et al. (2008). These authors found that
noticeable changes in the spectra occur at the early, no-flaring, stage of
emergence, whereas the begining of flare activity is accompanied by the
stabilization of the spectrum. This result also indicates relevance of the
multifractal organization of the photospheric magnetic fields to the flaring
activity.

5. As the resolution improves, the multifractality at small scales becomes more
pronounced: Hinode SOT/SP magnetograms spectra become increasingly steeper as
the spatial scale decreases. Thus, observations of ARs of the 24rd solar cycle
with high resolution instruments, such as Hinode SOT, SDO/HMI, and Big Bear
Solar Observatory New Solar Telescope (BBSO/NST, Goode et al. 2010) will bring
new opportunities to study solar magnetism via the multifractality approach.

The intermittent/multifractal nature of the magnetic field in solar active
regions attributes not {\it only} to the geometrical organization of the
solar-surface magnetism as it might be assumed from the first sight. Indeed, on
the contrary to a Gaussian process (where chaos is determined by a {\it sum} of
numerous independent variables, and high-order statistical moments are fully
determined by the first two moments), an intermittent process is rather a {\it
product} of large number of independent variables when high statistical moments
growth arbitrary with the number of the moment. These considerations, when
converted to physics, imply that random strong peaks in the solution of the
problem of the vector-field's transport by random flows (say, the magnetic field
vector in a turbulent electro-conductive flow) correspond to structural features
such as flux tubes and/or thin sheets of magnetic field lines. Strong
intermittency observed in complex ARs is a hint that we observe a
photospheric imprint of enhanced sub-photospheric dynamics.

Authors are thankful to Roberto Bruno, Vincenco Carbone, Manolis Georgoulis,
Philip Goode, James MsAteer, Jean-Claude Vial and Vadim Uritsky  for helpful
discussions and encouragement of this study. We are grateful to anonymous
referee whose criticism and comments helped to improve the paper. SOHO is a
project of international cooperation between ESA and NASA. Hinode is a Japanese
mission developed and launched by ISAS/JAXA, collaborating with NAOJ as a
domestic partner, NASA and STFC (UK) as international partners. Scientific
operation of the Hinode mission is conducted by the Hinode science team
organized at ISAS/JAXA. This team mainly consists of scientists from institutes
in the partner countries. Support for the post-launch operation is provided by
JAXA and NAOJ (Japan), STFC (U.K.), NASA (U.S.A.), ESA, and NSC (Norway). Hinode
SOT/SP inversions were conducted at NCAR under the framework of the Community
Spectro-polarimtetric Analysis Center (CSAC; $http://www.csac.hao.ucar.edu/$).
This work was supported by NSF grant ATM-0716512 and NASA LWS grant NNX08AQ89G.

{}

\end{document}